# Enhancement of Voltage, Ion current and Neutron Yield in Pyroelectric Accelerators

V. Sandomirsky, A. V. Butenko, and Y. Schlesinger[*]
Department of Physics, Bar-Ilan University, Ramat-Gan 52900, Israel
R. Levin
Department of Chemical Technology and Biotechnology, The J&S University Center,
Ariel 40700, Israel

Abstract.

Utilization of current pyroelectric accelerators (PEA) is limited due to low ion current and neutron generation yields. Current design, using pyroelectrics (PE) with high pyrocoefficient ($p$), having high dielectric constant ($\varepsilon$), limits the figure-of-merit. We present detailed analysis of a modified structure of PEA, providing the highest attainable voltage and ion current. In the paired configuration, using metal plates covering the polar faces, with grounded back plates, the accelerating voltage and electric field are proportional to $p$ and do not depend on $\varepsilon$. Therefore, in the modified structure, PE with high $p$ significantly increases the ion and neutron yields.

---

[*] Corresponding author: schlesy@mail.biu.ac.il

***The characteristics of the current PEAs.*** PEAs attract attention [1-5], due to their simple construction, palm-size dimensions and low cost. Commercial PE X-ray source is available and laboratory PE neutron generators are investigated. The yield attained ~ $10^5$ n/heating-cooling cycle. For the last 4 years the yield has been increased by an order of magnitude [3]. To become a viable technology, the yield must be increased by at least one more order of magnitude.

The basic element of PEA is a PE (usually LiTaO$_3$, with $p=2\times10^{-2}$ $\mu$C/cm$^2$K is used) cylinder, having a height and diameter of ~ 1 cm. The polarization is directed along its axis, perpendicular to faces. The PE temperature is raised by $\Delta T \approx$ 100 K, and the PE-effect creates a strong depolarizing electric field (DEF) ~$10^5$ V/cm in the PE vicinity. A metallic tip ~ 70 nm on the PE polar face enhances the local DEF up to ~$10^8$ V/cm to ionize the ambient D$_2$ gas. The D$^+$ ions are accelerated by DEF up to $\approx$ 200 keV, striking a deuterated conducting target placed opposite the PE face at a distance of ~1 cm. As a result, neutrons generate via the *D-D* reaction. The heating-cooling cycle lasts ~ 10 min. Practically, the neutrons are generated within ~ 100 s during the cooling. A schematic drawing of PEA used in [3] is shown in Fig. 1a.

***Electrostatic analysis of the present PEA.*** Consider PEA shown in Fig. 1a [3]. We calculate the voltage and the electric field in the air gap. This derivation is straightforward but often wrongly formulated. The surface charge on the metal layers at $z = L_1$ at any moment ($t$) is

$$Q(L_1,t) = Q_1(0,t) + Q_{2t}(L_1 + L_2, t), \quad (1)$$

where $Q_1$ and $Q_{2t}$ are the charges on the plane $z = 0$, and on the target plane $z = L_1 + L_2$, respectively. At $t = 0+$, after application of a temperature step $\Delta T$ to PE,

$$Q(L_1, 0+) = Sp\Delta T, \tag{1.1}$$

where $S$ is the area of metal plate.

The potential difference between the grounded planes at $z = 0$ and at $z = L_1 + L_2$ is 0. Hence, the voltages $U_P$ across PE and across the air gap ($U_g$) are

$$U_P(t) = U_g(t) = \frac{Q_1(t)}{C_1} = \frac{Q_{2t}(t)}{C_2}. \tag{2}$$

The capacitances of PE ($C_1$) and of the air gap ($C_2$) are

$$C_1 = \frac{\varepsilon S}{4\pi L_1}; \quad C_2 = \frac{S}{4\pi L_2}. \tag{3}$$

From (1), (2) and (3) we have

$$U_g(t) = \frac{4\pi L_2 Q(t)/S}{1+\varepsilon}; \quad E_g(t) = \frac{4\pi Q(t)/S}{1+\varepsilon}, \tag{4}$$

where $E_g$ is DEF in the air gap. At $t = 0+$

$$U_g(0+) = \frac{4\pi L_2 p\Delta T}{1+\varepsilon}; \quad E_g(0+) = \frac{4\pi p\Delta T}{1+\varepsilon} \tag{4.1}$$

Thus, the factors $pL_2/(1+\varepsilon)$ and $p/(1+\varepsilon)$ determine the potential difference and electric field in the presently used structure of PEA. *Therefore it is not advantageous using PEs with high p, since they usually have also a large ε.* Notice also that $U_g$ increases with $L_2$, as long as the parallel plate capacitor approximation is valid.

***Electrostatic analysis of the modified PEA.*** Consider the structure drawn in Fig. 1b, similar to those described in [2] and [4] (see also [6]).

Obviously, the charges on the metal sheets (at $z=0$ and $z=2L_1+L_2$) are

$$Q_1(0,t) = Q_3(2L_1 + L_2, t) = 0 \tag{5.1}$$

and

$$Q(L_1,t) = -Q_2(L_1+L_2,t); \quad Q(L_1,0+) = -Q_2(L_1+L_2,0+) = Sp\Delta T \quad (5.2)$$

Then, the voltage over the air gap and the field within it are

$$U_g(t) = 4\pi L_2 Q(t)/S; \; E_g(t) = 4\pi Q(t)/S;$$

$$U_g(0+) = 4\pi L_2 p\Delta T; \; E_g(0+) = 4\pi p\Delta T \quad (5.3)$$

Thus, <u>in this case</u> the voltage and the field are determined by the factors $pL_2$ and $p$, both <u>not depending</u> on $\varepsilon$. Hence, one can increase both values substantially, choosing PE with high $p$. The advantage of the structure in Fig. 1b is evident.

***Estimate of upper limit of neutron gain.*** In the following, we estimate the maximal possible ion yield and neutron gain in the structures depicted in Fig. 1a and 1b.

Let us consider the relaxation of the surfaces charges on PEs and on the metal plates, and, respectively, the DEF relaxation. We will refer to Fig. 1a, but the same arguments hold also for Fig. 1b.

It is convenient to divide the surface charges on metal layers into two parts. For example, the polarization charge at $z = L_1$ is $Q(L_1,t) > 0$. Then the charge, screening the polarization charge on the metal/ PE interface, is $Q_{mi}(L_1,t) = -Q(L_1,t)$, and the charge on the metal/air interface is $Q_{me}(L_1,t) > 0$. Accordingly, at $t = 0+$

$$Q(L_1,0+) = -Q_{mi}(L_1,0+) = Q_{me}(L_1,0+) = Sp\Delta T \quad (6)$$

Thus, the left metal surface at the air gap acquires a positive charge equal to the polarization charge of PE.

As PEs are protected from adsorption screening by the metal plates (the screening by internal charge carriers is much slower), the polarization charges relax only due to cooling decreasing $p\Delta T(t)$. However, the charge $Q_{me}(L_1,t)$ can relax also due to the adsorption. Assume there is no adsorption present. Then, if $\Delta T(\delta t) = \Delta T(0+) - \delta T$, so

$$Q(L_1, \delta t) = Q(L_1, 0+) - \delta Q \tag{7.1}$$

Respectively,

$$Q_{mi}(L_1, \delta t) = -[Q(L_1, 0+) - \delta Q]; \quad Q_{me}(L_1, \delta t) = Q(L_1, 0+) - \delta Q \tag{7.2}$$

Thus, the charge $-\delta Q$ from the metal/PE side and the charge $\delta Q$ from the metal/air side "recombine" inside the metal. The charge $-\delta Q$ splits between the metal target layer ($z = L_1 + L_2$) and the external metal plate $z = 0$, into two parts, $\delta Q_2/\delta Q_1 = C_2/C_1$ (see Eq. 2). This rearrangement proceeds through the common ground. The metal/PE charges at $z = 0$ are also described by Eq. (7.1).

Let us now consider the opposite situation, when cooling is absent, and $Q(L_1, t) = -Q_{mi}(L_1, t) = Sp\Delta T(0+) = const$. The charge $Q_{me}(L_1, t)$ relaxes only due to adsorption. It means that in the air gap appear two particles with opposite charges $\delta q^+ = -\delta q^-$. The $\delta q^-$ particle attracts to $z = L_1$, and $Q_{me}(L_1) \to Q_{me}(L_1) + \delta q^-$, and the $\delta q^+$ particle attracts toward the target at $z = L_1 + L_2$. There its charge divides between the target $(\delta q_2^+)$ and the outward metal layer at $z = 0$ $(\delta q_1^+)$, as $C_2/C_1$. So,

$$Q_{2t} \to Q_{2t} + \delta q_2^+ \ (Q_{2t} < 0) \text{ and } Q_{me}(0) \to Q_{me}(0) + \delta q_1^+ \ (Q_{me} < 0).$$

Thus, the polarization charge and the compensating charge on the inside metal/PE interface do not change. Only the charges on the exposed to air sides of the metal plates relax. This relaxation route provides the ion current impinging on the target. The maximal ion charge, crossing the air gap, capable to initiate a nuclear reaction is

$$Q_{\max 1} \leq Sp\Delta T(0+) - U_{th}(C_2 + C_1), \tag{8}$$

where $U_{th}$ is the threshold voltage for the nuclear reaction. The sign " < " takes into account the losses due to cooling and adsorption of some "parasitic" charged particles.

For the case of Fig. 1b

$$Q_{\max 2} \leq Sp\Delta T(0+) - U_{th}C_2 \qquad (9)$$

The above analysis is based on the validity of the parallel-plate capacitor approximation. Hence, the lateral dimensions of the metal plates must be larger then the gap, which can be easily satisfied. The equipotential metal plate, covering the total PE face, maximizes the yield of the accelerated ions in comparison with the bare, non-equipotential, PE face.

Eqs. (8) and (9) allow estimating the maximum neutron gains attainable in [3] and [4]. The structure in [3] is closely similar to that shown in Fig. 1a [7]. One has

$$p = 2 \times 10^{-8}\,\text{C/cm}^2\text{K}\,; \varepsilon = 43\,; \Delta T = 100\,\text{K}\,; S = 7.06\,\text{cm}^2\,; L_1 = L_2 = 1\,\text{cm}.$$

Assuming $U_{th} \simeq 80\,\text{kV}$, then $U_{g1}(0+) = 514$ kV; $E_{g1}(0+) = 514$ kV/cm, and

$$Sp\Delta T(0+) - U_{th}(C_2 + C_1) = 1.21 \times 10^{-5}\,\text{C} = 7.56 \times 10^{13} \text{ single-charged ions} \qquad (10.1)$$

If these ions flow during 100 s, the average ion current is 121 nA. Ref. [3] does not provide the product of the number density of deuterium ions in the target × total cross-section of the *D–D* reaction for the target thickness used. Ref. [4] cites this value as $\rho_t \Sigma = 3.03 \times 10^{-8}$. Assuming this value for both cases, the upper theoretical limit of the gain and the ratio of experimental and the maximum gains are

$$7.56 \times 10^{13} \times 3.03 \times 10^{-8} = 2.3 \times 10^6 \text{ n}\,; \frac{1.9 \times 10^5}{2.3 \times 10^6} = 8\% \qquad (10.2)$$

Let us do the same estimates for the structure in Fig. 1b. The structure described in [2] and [4] is close to that shown in Fig. 1b (see also note [6]). All data are as given above, except $S = 3.14\,\text{cm}^2$. Then, according to (9)

$$U_{g2}(0+) = 4\pi L_2 p \Delta T(0+) = 22.6\,\text{MV};$$
$$E_{g2}(0+) = 4\pi p \Delta T(0+) = 22.6\frac{\text{MV}}{\text{cm}} \qquad (12)$$

$$S p \Delta T(0+) - U_{th} C_2 = 6.2 \times 10^{-6}\,\text{C} = 3.88 \times 10^{13} \text{ single-charged ions} \qquad (12.1)$$

If this ion flow lasts for 100 s, then the average ion current is 62 nA. The maximal theoretical gain and the ratio of experimental and the maximum gains are

$$3.88 \times 10^{13} \times 3.03 \times 10^{-8} = 1.18 \times 10^6 \text{ n;} \quad \frac{5.9 \times 10^4}{1.18 \times 10^6} = 5\% \qquad (12.2)$$

Thus, the modified structure showed in Fig. 1b with PE having a larger $p$, one could increase substantially the neutron gain, and possibly initiate reactions with higher threshold energy [8].

The analysis conducted shows also that to increase the ion gain in the structures shown in Fig. 1a and Fig. 1b, it is advantageous to keep the temperature step $\Delta T(0+)$ fixed, and preventing the cooling of PE.

6. The configuration in Fig 1b is similar, but essentially not identical, to those described in [2], [4]. There, *only a part of* the PE face at $z = L_1$ is metallized, resulting in a decreased gain.

7. It is possible, however, that in [3] the flat capacitor approximation was not valid.

8. Both theoretical estimates (10.2) and (12.2), and the ratios (11) and (13), may be approximate, since they are based on the value of $\rho_t \Sigma = 3.03 \times 10^{-8}$ from [2]. The latter suggested monoenergetic flow of *D*-ions with energy of 100 keV. This may be not valid for the energies assumed in the above analysis. This, however, does not affect the final conclusions.

**Figure captions**

Fig. 1 (a) Single PE metal covered configuration

(b) Twin PE metal covered configuration

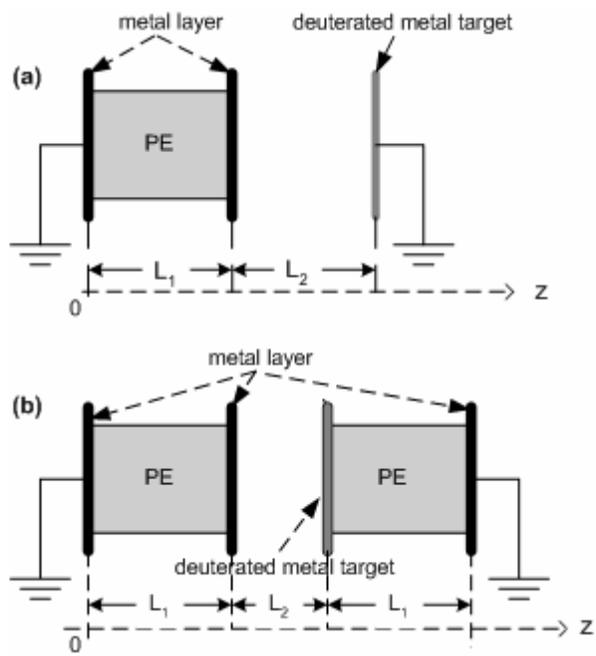

Fig. 1  (a) Single PE metal covered configuration

(b) Twin PE metal covered configuration